# Evolution of clusters in energetic heavy ion bombarded amorphous graphite-III: Photoemission and velocity analysis of sputtered carbon clusters ($\boldsymbol{C_m^{0,\pm}, m < 4}$)

A. Qayyum[1], M. N. Akhtar[1], T. Riffat[1], and Shoaib Ahmad[2,*]
[1]*Accelerator Laboratory, PINSTECH, P. O. Nilore, Islamabad, Pakistan*
[2]*National Centre for Physics, QAU Campus, Shahdara Valley Road, Islamabad 44000, Pakistan*

[*]Email: sahmad.ncp@gmail.com

## Abstract

Diagnostics of the carbon atoms, ions, and clusters $C_m^{0,\pm}$ ($m < 4$) sputtered from a graphite surface under 10 keV $Xe^+$ bombardments reveals the presence of neutral $C_m^0$ and positively charged $C_m^+$ species in the photoemission spectra. The observed absence of $C_m^+$ in the mass spectra by a velocity analyzer is supplemented by the presence of the negatively charged $C_m^-$. The velocity spectra are dominated by $C_1^-$ and $C_2^-$. $C_3^-$ and $C_4^-$ are also observed having much reduced peak intensities. These results may help us to understand the contribution of neutral and charged species in the heavy-ion sputtering of graphite and the energetics of soot formation. $Xe^+$ is chosen being the heaviest noble gas ion irradiating at not too high energy i.e 10 keV.

Excited and charged sputtered carbon species $C_m^{0,\pm}$ ($m < 4$) have been produced by 10 keV $Xe^+$ beam incident on graphite surfaces. The mass spectrometry data on the neutral and charged sputtered species is likely to illustrate the recently observed heavy carbon clusters and closed cages from similar irradiations but at higher energies [1]. These were investigated by the twin techniques of secondary ion mass spectrometry and ion-induced photoemission spectroscopy [2] of the sputtered species. The motivation for the simultaneous detection comes from the need to know the cluster status of the sputtered carbon species ~i.e., $m \geq 2$ as well as the relative constitution of the excited and the charged states of the respective species at the point of ion impact.

Diagnostics of the sputtered species is done by varying the ion incidence angle a with respect to the surface normal. The $Xe^+$ beam energy was also varied but we have chosen to present the spectra for the fixed energy, i.e., 10 keV with 10 µA current, as at that energy the beam sputters ~1.5 $\pm$ 0.2 $[C/ion]$ [3]. This sets the lower limit on the beam energy for the twin diagnostics, especially the mass analysis. A monochromator analyzes the spectrum of the ion-induced photons emitted by the sputtered excited and charged species. We have used an indigenously developed velocity filter for the mass analysis of the charged sputtered species. The velocity filter is described in detail elsewhere [4]. Since the energy of these clusters is very small $\sim E_b/2$, where $E_b/2$ is the average surface binding energy of the atoms and clusters. In the experiments reported here, we biased the target with voltage $V_t$ equal to $\pm 1$ kV with respect to the extraction aperture to impart a well-defined energy and to extract the corresponding charged ions.

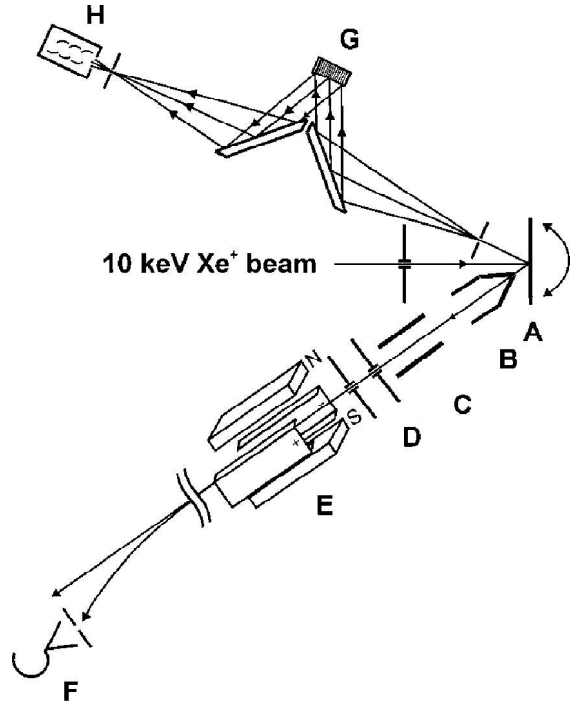

FIG. 1. Schematic diagram of the experiment is shown with the normally incident Xe$^+$ ion beam, vertically rotatable target (A) to vary the angle $\alpha$. An electrostatic extraction (B) and focusing lens (C), collimating apertures (D), the velocity filter (E), and a channel electron multiplier (F) are part of the mass diagnostic. A monochromator with wavelength selection (G) and photon detection (H) is fitted at 45° to the beam.

The schematics of the experimental setup are shown in Fig. 1. The $Xe^+$ beam is incident on the target and the monochromator is positioned to look at the point of ion impact through a fused silica window. Similarly, the extraction aperture for charged ions is placed at right angles to the line of sight of the monochromator. At a $\alpha = 0°$, both the detection instruments are symmetrically placed around the beam direction. At other angles, the target is rotated around a vertical axis towards one instrument or the other. The pressure in the target chamber is maintained with an ion pump at $\sim 10^{-8}$ mbar; during ion bombardment it rose to $\sim 10^{-6}$ mbar. The collimating apertures for the velocity filter restrict the divergence of the extracted beam to $\pm 0.2°$. Charged-particle detection is done by a channel electron multiplier (CEM).

Figure 2 shows the spectrum of light emitted by the sputtered species at $\alpha = 45°$. The range of the chosen wavelength is between 180 and 600 nm to be able to detect the excited atomic carbon $C_1$ line (CI) at 193 nm and to record the other lines, and possibly the bands due to the clusters $C_m (m > 2)$ within this range.

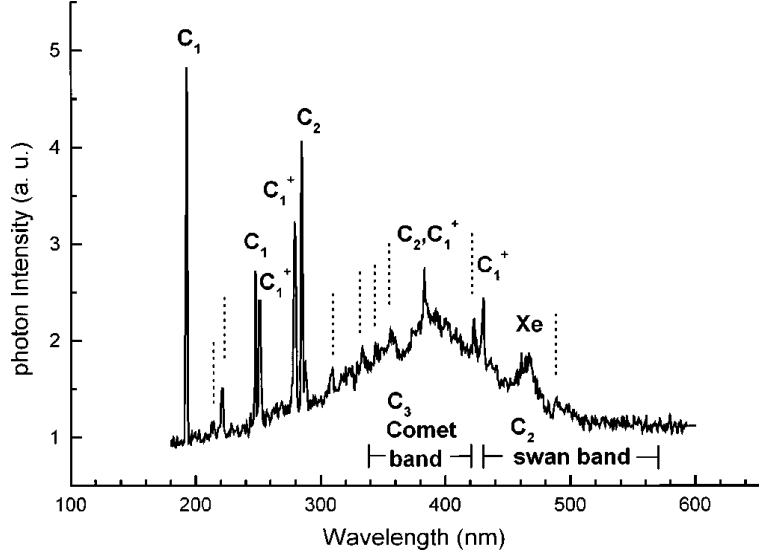

FIG. 2. Photoemission spectrum of light emitted by the sputtered species at $\alpha = 45°$ is shown for the range of wavelength 180-600 nm. Notice the broad hump between 250 and 500 nm.

The characteristic lines due to $C_1$, $C_1^+$, $C_1^{++}$, and the excited diatomic molecule $C_2$ are superimposed on a broad hump between 250 and 500 nm. The *Xe* lines are also noticeable. In the case of $C_1$, we see a massive peak at 193 and a second one at 248 nm. The presence of singly and doubly charged $C_1^+$, $C_1^{++}$ is confirmed in the sputtered species on or just outside the surface. The first cluster to be positively identified is $C_2$ with a large peak at 285 nm. Higher clusters with *m*>2 are not unambiguously identifiable since the band heads associated with them are fairly weak in contrast to the emission lines. However, the weak presence of these bands is marked in Fig. 2 by the appropriate band names. The emission lines and bands are interpreted using Reference [5-7].

The other important feature of the spectrum shown in Fig. 2 is the broad hump between 250 and 500 nm. This practically covers the entire measured range and is peaked around 380 nm. There are various unidentified weak peaks as well as broader band-like features that we have marked with vertical dotted lines on the spectrum in Fig. 2 that may be due to: (1) the overlap of the adjacent peaks in this range that are not completely resolvable due to the inherent resolution of 1 nm of the monochromator, (2) the cluster band overlap for the $C_m (m \geq 2)$, or (3) the light emission from the surface plasmons in the corresponding energy range 2.5–5 eV [8].

The unambiguous identification of clusters comes from the negatively charged $C_m^-$ $(\geq 2)$ mass spectrum that is always dominated by the $C_1^-$ peak. Such a spectrum at a $\alpha = 45°$ is shown in Fig. 3. The spectrum shows the corresponding peaks due to $C_1^-$, $C_2^-$, $C_3^-$, and $C_4^-$. The intensities of the $C_3^-$ and $C_4^-$ peaks are shown five times enhanced and are about an order of magnitude lower than those due to $C_1^-$ and $C_2^-$. The other significance of mass analysis is the complete absence of the positively charged species $C_m^+$ in the +ve spectra.

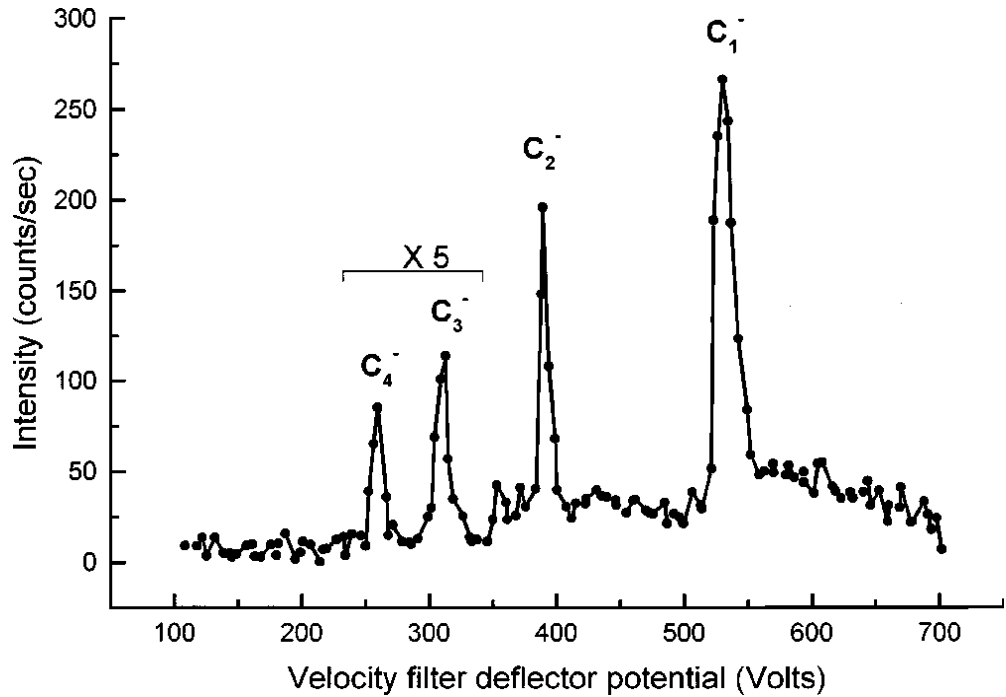

FIG. 3. Mass spectrum obtained with the ExB velocity filter at $\alpha = 45°$ is shown identifying the peaks due to $C_1^-$, $C_2^-$, $C_3^-$, and $C_4^-$; the latter two peaks are shown as multiples of 5.

While we have detected strong peaks due to the singly as well as doubly charged positive ions $C_1^+$, $C_2^+$ in the photoemission spectra as shown in Fig. 2, these do not survive long enough to be mass analyzed. This is possibly due to neutralization of the sputtered ions.

The trend for the relative intensities of the sputtered species' characteristic lines has been plotted in Fig. 4 taken from photoemission data as a function of a. The line intensities are plotted for $C_1$, $C_1^+$, and $C_2$. The most prominent lines have been selected for the comparison of relative intensities. The intensity ratios for $C_1$ and $C_1^+$ do not show a trend that follows the pattern of variation of sputtering yield with the incidence angle $\alpha$; the yield $\propto 1/cos\alpha$. The dotted trend line, predicted by linear cascade sputtering theory [9], is also shown in Fig. 4. The linear cascade sputtering theory is mostly valid for medium mass ion bombardment of heavy-metal targets.

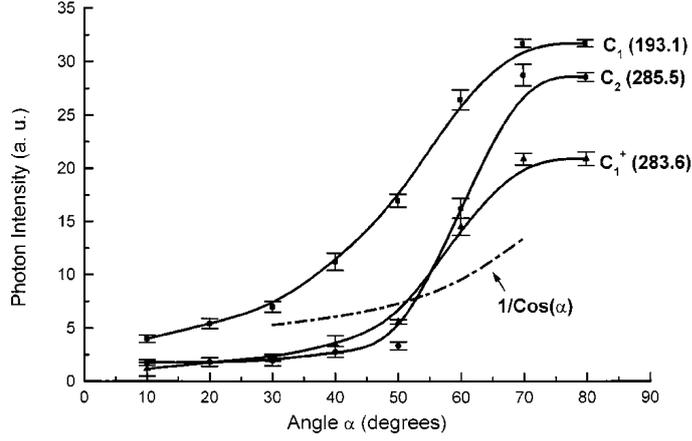

FIG. 4. Trend for the relative line intensities of $C_1$, $C_1^+$, and $C_2$ has been plotted from the photoemission data as a function of a. The 1/cos $\alpha$ curve is also shown for comparison with the variation of sputtering yield with ion incidence angle.

The light and heavy ions introduce nonlinearities in the sputtering yields and emission properties [10]. The observed disagreement of $C_1$ and $C_1^+$ intensity variation with angle a shows that the linear cascade sputtering theory is not strictly valid for the $Xe^+ \rightarrow C$ system. Also, the $C_2$ intensities do not fit the sputtering yield pattern and show a much sharper increase after $\alpha \approx 50°$. We may interpret it as the combined effect of the bulk sputtering along with that of the reconstituted surface with freshly adsorbed carbon layers—the regenerative soot [11]. The enhanced yield points to the reduction in binding forces of the surface constituents. At large a, higher sputtering yields imply surface coverage with loose agglomeration of $C_1$ forming a fresh-soothed surface denominated by $C_2$ and possibly higher clusters. Therefore, the yield of the $C_1$ is unaffected, but $C_2$ shows a marked increase.

The presence of neutral and positively charged $C_m^{0,+}$ in the emitted line spectra of the sputtered carbon species close to the surface and the subsequent absence of $C_m^+$ ($m \geq 1$) from the mass spectra identifies their high electron attachment probability. Our data suggest that the twin processes

(1) $C_m^+ + e \rightarrow C_m^0$,
(2) (2) $C_m + e \rightarrow C_m^-$

dominate the sputtered species' neutralization and electron attachment. These results help us to understand the well-noted absence of the $C_1^+$ peak from the direct recoil (DR) spectra in our earlier work [1]. As regards the larger clusters with $m > 4$, we have not detected these with certainty in the mass analysis nor can we identify the predicted [7] emission lines and bands belonging to larger clusters with $3 < m < 9$ in the photoemission spectra. Further work is in progress.